\documentclass[11pt]{llncs}

\usepackage{amsmath}
\usepackage{amssymb}
\newcommand{\Prf}{\proof}

\newcommand{\bra}[1]{\langle #1 |}
\newcommand{\ket}[1]{| #1 \rangle}

\newcommand{\ketbra}[2]{\ket{#1}\bra{#2}}
\newcommand{\proj}[1]{\ket{#1}\bra{#1}}
\newcommand{\tr}{{\rm Tr}}
\newcommand{\one}{{\bf 1}}

\newcommand{\Hil}{\mathcal H}
\newcommand{\chan}{\mathcal N}
\newcommand{\cchan}{\widehat {\mathcal N}}
\newcommand{\chanh}{\mathcal N^\dagger}
\newcommand{\cchanh}{\widehat {\mathcal N}^{\dagger}}

\newcommand{\cref}[1]{(\ref{#1})}

\newcommand{\zeroset}[1]{{\mathbb K}}

\newcommand{\half}{\frac{1}{2}}
\newcommand{\id}{{\rm id}}

\begin{document}

\title{Conditions for the approximate correction of algebras}
\author{C\'edric B\'eny}
\institute{Centre for Quantum Technologies, National University of Singapore}
\maketitle

{\abstract 
We study the approximate correctability of general algebras of observables, which represent hybrid quantum-classical information. This includes approximate quantum error correcting codes and subsystems codes. We show that the main result of \cite{kretschmann08} yields a natural generalization of the Knill-Laflamme conditions in the form of a dimension independent estimate of the optimal reconstruction error for a given encoding, measured using the trace-norm distance to a noiseless channel.
}



Slightly relaxing the requirement of perfect quantum error correction can allow for significantly larger quantum codes~\cite{leung97,crepeau05}. Here we focus on a quantification of the correction error based on the diamond norm distance, introduced below, which can be related to the worst case entanglement fidelity. (See~\cite{barnum00} for the case of average entanglement fidelity). There exists results giving sufficient conditions for a code to be approximately correctable in that sense~\cite{schumacher01,doddamane09}, however it is not known how general these conditions are. Instead we want to draw attention to results by Kretschmann {\it et al.}~\cite{kretschmann08,kretschmann08x1} who gave lower and upper bounds for the optimal reconstruction error for a given code in terms of the complementary channel's distance to a maximally forgetful channel. The present report can be seen partly as an advertisement of these results in a context where they are not widely known, or their meaning not recognized, namely as a providing a necessary and sufficient condition for approximate error correction. In addition we improve on these results by rendering the conditions more explicit, and generalizing them to the correction of general algebras.


The condition that we obtain (Theorem~\ref{mainthm}) can be understood as a perturbation of the exact Knill-Laflamme condition~\cite{knill97}, or more generally its subsystem version~\cite{kribs05}, or full algebraic form~\cite{beny07x1}. We also give an essentially equivalent condition based on individual observables of the algebra (Theorem~\ref{secthm}). The correctable algebra can be understood as representing a quantum system with superselection rules, or a hybrid quantum-classical memory~\cite{kuperberg02}, and can be shown to be the most general type of exactly correctable information in the sense of~\cite{blume-kohout07}. 





\section{Preliminaries}

 
A channel $\chan$ is a completely positive trace-preserving map. It can always be written as
\[
\chan(\rho) = \sum_i E_i \rho E_i^\dagger
\]
where the operators $E_i$ are the channel elements and must only satisfy $\sum_i E_i^\dagger E_i = \one$. 
The dual $\chan^\dagger$ is defined by the relation
\[
\tr(\chan(\rho) A ) = \tr(\rho \chan^\dagger (A))
\]
for any state $\rho$ and any operator $A$. This implies that 
\[
\chanh(A) = \sum_i E_i^\dagger \rho E_i.
\]
Physically, $\chan$ is interpreted as evolving states, while $\chan^\dagger$ evolves observables. Hence $\chan^\dagger$ represents the Heisenberg picture for the evolution defined by the channel. To avoid confusion, we only call $\chan$ a {\em channel}, while $\chan^\dagger$ is its {\em dual}. 

\subsection{Complementary channel}

For any channel $\chan$ we can find an isometry $V$ ($V^\dagger V = \one$) such that
\[
\chanh(A) = V^\dagger (A \otimes \one) V.
\]
The isometry $V$ amounts to adding an extra system; the ``environment'', with a fixed pure initial state $\ket{\phi_E}$ and letting it interact unitarily with the system for a fixed amount of time, i.e. $V \ket{\psi} := U (\ket{\psi} \otimes \ket{\phi_E})$ for some unitary operator $U$.
This allows one to define a {\em complementary channel} $\cchan$ through
\[
\cchanh(B) =  V^\dagger (\one \otimes B) V.
\] 
The channel $\cchan$ maps the initial state of the system to the final state of the environment.
The most important fact that we will use is that all complementary channels are equivalent up to a unitary transformation of their output, and eventual embedding into a larger environment, and that this property is stable under perturbation as shown in~\cite{kretschmann08}.

It is easy to relate a dilation with isometry $V$ to channel elements $E_i$ by introducing any orthonormal basis $\ket{i}$ of the environment as follows:
\[
\chanh(A) = V^\dagger (A \otimes \one) V = \sum_i V^\dagger (A \otimes \proj{i}) V = \sum_i V^\dagger (\one \otimes \ket{i}) A  (\one \otimes \bra{i}) V
\]
Hence we can use
\[
E_i = (\one \otimes \bra{i}) V
\]
which is defined by
\[
\bra{\psi} E_i = (\bra{\psi} \otimes \bra{i}) V.
\]
This implies that the complementary channel can be written in dual form as
\[
\cchanh(B) = \sum_{ij} \bra{i} B \ket{j} E_i^\dagger E_j.
\]

\subsection{Distance between channels}
\label{distance}
 
Any operator $A$ has a norm defined by
\[
\|A\| := \sup_{\ket{\psi}} \frac{\|A \ket{\psi} \|}{\|\ket{\psi}\|}.
\]
This norm on operators can be used to define a distance between dual channels as follows:
\[
\| \chanh_1 - \chanh_2 \| := \sup_{A, \|A\| \le 1} \|\chanh_1(A) - \chanh_2(A)\|.
\]
However this distance can increase when the channels are tensored with the identity channel on an auxiliary space. Therefore we also define the {\em completely bounded} norm
\[
\| \chanh_1 - \chanh_2 \|_{cb} := \| (\chanh_1 - \chanh_2) \otimes \id \|
\]
where $\id$ is the identity channel on a Hilbert space of the same dimension as that of the source of the two channels. This distance is guaranteed to be stable under further trivial extension (See~\cite{johnston07} for an introduction). It is equal to the {\em diamond norm} distance between the channels themselves:
\[
\| \chanh_1 - \chanh_2 \|_{cb} = \| \chan_1 - \chan_2 \|_{\diamond}
\]
which is defined by
\[
\begin{split}
\| \chan_1 - \chan_2 \|_{\diamond} &:= \| (\chanh_1 - \chanh_2) \otimes \id \|_1\\
&= \sup_{\rho} \tr |((\chan_1 - \chan_2) \otimes \id) (\rho)|.
\end{split}
\]
where $\|\cdot \|_1$ is the trace norm.   
This distance is directly related to the worst case probability of failing to distinguish between
the outputs of the two channels for any common initial state. 






\section{Exact correctability of algebras}

We review here results on the exact correctability of algebras, as defined in \cite{beny07x1}. 
A $\dagger$-algebra  (or algebra for short) is a set of operators closed under multiplication and which also contains the adjoint of all its elements. For instance, suppose that our Hilbert space $\Hil$ is divided into two subsystems: $\Hil = \Hil_A \otimes \Hil_B$, then consider the set $\mathcal A$ of operators of the form $A \otimes \one$, where $A$ is an operator on $\Hil_A$ and $\one$ the identity on $\Hil_B$. It is trivial to show that $\mathcal A$ is an algebra. It represents all the local observables acting on $\Hil_1$. In fact this is close to being the most general form of a $\dagger$-algebra. For any $\dagger$-algebra $\mathcal A$ we can find a decomposition of the Hilbert space into orthogonal subspaces $\Hil_i$ which are left invariant by all elements of the algebra. Furthermore, when restricted to any of these invariant subspaces, the algebra has precisely the form described in the above example. Hence the algebra defines a set of subsystems living in a family of orthogonal subspaces. This means that any element $A \in \mathcal A$ is of the form
\[
A = \sum_i A_i \otimes \one_i
\]
where $A_i \otimes \one_i$ is an operator supported on $\Hil_i$. Said differently, if $P_i$ is the projector on $\Hil_i$ then $P_i A P_i = A_i \otimes \one_i$. 

A useful tool that we will be using is the projector $\mathcal P_{\mathcal A}$ on this algebra, which we take to be orthogonal in terms of the Hilbert-Schmidt inner product between operators. This is a quantum channel satisfying $\mathcal P_{\mathcal A}^2 = \mathcal P_{\mathcal A} = \mathcal P_{\mathcal A}^\dagger$, whose range is precisely $\mathcal A$. It has the following explicit form:
\begin{equation}
\label{proj}
\mathcal P_{\mathcal A}(\rho) = \sum_i \frac{1}{\tr P_i}\sum_{jk} (\one_i \otimes \ket{j}_i\bra{k}_i ) \; \rho \; (\one_i \otimes \ket{k}_i\bra{j}_i )
\end{equation}
where the vectors $\ket{j}_i$ for a fixed $i$ are orthogonal and satisfies $\sum_k \one_i \otimes \ket{k}_i\bra{k}_i = P_i$.

We say that an algebra $\mathcal A$ is correctable for the channel $\chan$ if there exists a ``correction'' channel $\mathcal R$ such that for all $A \in \mathcal A$,
\begin{equation}
\label{ecd}
(\mathcal R \circ \chan)^\dagger(A) = A.
\end{equation}


Note that $\mathcal A$ contains the spectral projectors of any observable $A \in \mathcal A$. Hence this definition implies that measuring $A$ before the action of the channel $\chan$ or after the correction will yield the same probabilities, no matter what the initial state was.

Clearly, Equ.~\ref{ecd} implies that $\mathcal P_{\mathcal A}\circ \mathcal R \circ \chan = \mathcal P_{\mathcal A}$. 
Hence an equivalent formulation is to require the existence of a (possibly different) channel $\mathcal R$ such that
\begin{equation}
\label{ecd2}
\mathcal R \circ \chan = \mathcal P_{\mathcal A}. 
\end{equation}

 
It was shown in~\cite{beny07x1} that any algebra $\mathcal A$ is correctable if and only if its elements $A \in \mathcal A$ all satisfy
\begin{equation}
\label{gkl}
[A, E_i^\dagger E_j ] = 0 \quad \text{ for all $i,j$}
\end{equation}
where $E_i$ are the error operators, or elements of the channel $\chan$ representing the interaction with the environment. 
What this means is that all the correctable algebras belong to the largest correctable algebra defined by the set of all operators commuting with the operators $E_i^\dagger E_j$, which is always a $\dagger$-algebra. 
This condition shows that the correctability of an algebra is conditioned purely on the correctability of a family of generators (for instance two different Pauli operators if we are correcting a qubit). 

We say that an observable $A$ is correctable if the algebra ${\rm Alg}(A)$ it generates is correctable. ${\rm Alg}(A)$ is commutative and spanned by the spectral projectors of $A$. Any other observable in that algebra is just a coarse-graining of $A$. 
Clearly all the correctable observables are correctable by the same correction channel, namely the one correcting the full commutant of the operators $E_i^\dagger E_j$.



Equation \ref{gkl} has a clear physical meaning if we note that the operators $E_i^\dagger E_j$ come from the complementary channel $\cchanh$. Indeed,
\[
E_i^\dagger E_j = \cchanh(\ketbra{i}{j}).
\]
Hence Equ. \ref{gkl} can also be written as
\begin{equation}
\label{gkl2}
[A,\cchanh(B)] = 0 \quad \text{ for all $B$.}
\end{equation}
The operators of the form $\cchanh(B)$ characterize the properties of the source system which are faithfully represented in the environment. Indeed, assuming that $\rho$ is any arbitrary state of the source system, if $B_i$ are elements of a POVM on the environment, then measuring $\{B_i\}$ yields probabilities $p_i = \tr(\cchan(\rho) B_i) = \tr( \rho \cchanh(B_i))$. These probabilities are precisely the probabilities that one would obtain by measuring the POVM with elements $A_i = \cchanh(B_i)$ on $\rho$. Hence the POVMs with elements of the form $\cchanh(B_i)$ for any POVM $\{B_i\}$ are observables of the source which represent information that is present in the environment. We say that these observables are {\em preserved}~\cite{beny08x1} by the channel $\cchan$. In particular the projective (sharp) observables (characterized by $A_i^2 = A_i$) which are preserved by a channel are the correctable observables for that channel~\cite{beny08x5}.

Hence Equ. \ref{gkl2} means that the correctable observables are precisely those which are compatible with the observables preserved in the environment. Theorem~\ref{secthm} below shows how this fact generalizes in the approximate case.

Let $\mathcal A'$ be the commutant of $\mathcal A$, i.e. the algebra formed by the operators which commute with all the operators of $\mathcal A$. If $\mathcal A$ is the correctable algebra, then $\mathcal A'$ is the algebra generated by the operators $\cchanh(B)$ for any $B$, or equivalently by the operators $E_i^\dagger E_j$ for any $i$ and any $j$. 
 
The correctability condition expressed in Equ. \ref{gkl} can also be written as
\begin{equation}
\label{gkl3}
\cchan = \cchan \circ \mathcal P_{\mathcal A'}.
\end{equation}
Indeed, this means that $\cchanh = \mathcal P_{\mathcal A'} \circ \cchanh$, and hence any operator in the range of $\cchanh$ commutes with all elements of $\mathcal A$. 
If $\mathcal P_{\mathcal A}$ is given by Equ.~\ref{proj} then one can show that $\mathcal P_{\mathcal A'}$ is given by
\begin{equation} 
\label{cproj}
\mathcal P_{\mathcal A'}(\rho) = \sum_i \frac{1}{\tr P_i} \sum_{jk} ( \widetilde {\ket{j}}_i\widetilde {\bra{k}}_i \otimes \one_i )\; \rho\; (\widetilde {\ket{k}}_i\widetilde {\bra{j}}_i \otimes \one_i ).
\end{equation}
where the vectors $\widetilde{\ket{j}}_i$ for a fixed $i$ are orthogonal and satisfy $\sum_k \widetilde{\ket{k}}_i\widetilde {\bra{k}}_i \otimes \one_i = P_i$.

Note that we have not mentioned any encoding, or code. The reason is that the encoding map can be considered to be included in the channel $\mathcal N$. 
For instance if the initial states are guaranteed to be encoded in a subspace $\Hil_C \subset \Hil$, i.e. the encoding is an isometry $V$ (i.e. such that $VV^\dagger$ projects on $\Hil_C$), then we immediately see by replacing the channel elements $E_i$ by $E_i V$ that an observable $A$ is correctable under this assumption if and only if
\[
[A, V^\dagger E_i^\dagger E_j V] = 0.
\]
If we require that the algebra formed by these operators is the whole algebra of operators on the code spans $\Hil_C$ then we recover the Knill-Laflamme conditions~\cite{knill97}, since this implies
\[
V^\dagger E_i^\dagger E_j V \propto \one_C.
\]
Similarly, if we only require the algebra to be that of all operators acting on a subsystem of the code we recover the conditions for subsystem error correction~\cite{kribs06}.


\section{Approximate correctability of algebras}
We will focus on the following approximate version of Equ. \ref{ecd2}:
\begin{definition}
\label{mydef}
We say that an algebra $\mathcal A$ is $\epsilon$-correctable for the noise channel $\chan$ if there exists a channel $\mathcal R$ such that
\[
\| \mathcal R \circ \chan - \mathcal P_{\mathcal A} \|_{\diamond} \le \epsilon.
\]
We define the minimal reconstruction error to be
\[
E_{\mathcal A}(\chan) := \min_{\mathcal R} \| \mathcal R \circ \chan - \mathcal P_{\mathcal A} \|_{\diamond}.
\]
\end{definition}

The following theorem gives a ``necessary'' and ``sufficient'' condition for approximate error correction of an algebra in the form of an estimate of the optimal correction error. 

\begin{theorem}
\label{mainthm}
Let
\begin{equation}
\label{estim}
\delta_{\mathcal A}(\chan) = \| \cchan - \cchan \circ \mathcal P_{\mathcal A'} \|_{\diamond} 
\end{equation}
then
\[
\frac{1}{4}\delta_{\mathcal A}^2 (\chan) \le E_{\mathcal A}(\chan) \le 2\delta_{\mathcal A}^{\half}(\chan).
\]
\end{theorem}
Note that $\delta_{\mathcal A}(\chan)$ is explicit apart from the diamond norm (see~\cite{watrous09} or~\cite{johnston07} for computation techniques). 
\Prf
These conditions follow from the exact condition (Equ.~\ref{gkl3}) and the main result of \cite{kretschmann08}, 
namely that if 
\[
\| \chan_1 - \chan_2 \|_{\diamond} \le \epsilon
\]
then for all channels $\cchan_1$ complementary to $\chan_1$ there exists a channel $\cchan_2$ complementary to $\chan_2$ such that
\[
\| \cchan_1 - \cchan_2 \|_{\diamond} \le 2 \sqrt{\epsilon}.
\]
Suppose that for some channel $\chan$,
\[
\| \cchan - \cchan \circ \mathcal P_{\mathcal A'} \|_{\diamond} \le \epsilon.
\]
We know from Equ.~\ref{gkl3} that the algebra $\mathcal A$ is correctable for any channel $\mathcal M$ complementary to $\cchan \circ \mathcal P_{\mathcal A'}$. In addition we can choose $\mathcal M$ such that
\[
\|\chan - \mathcal M \|_{\diamond} \le 2\sqrt{\epsilon}
\]
Let $\mathcal R$ be the correction channel for this choice of $\mathcal M$, i.e. $\mathcal R \circ \mathcal M = \mathcal P_{\mathcal A}$, then we have
\[
\|\mathcal R \circ \chan -  \mathcal P_{\mathcal A} \|_{\diamond} = \|\mathcal R \circ \chan - \mathcal R \circ \mathcal M\|_{\diamond} \le  \|\mathcal R\|_{\diamond} \|\chan - \mathcal M\|_{\diamond} \le 2\sqrt{\epsilon}.
\]

Reciprocally, suppose that  $\mathcal A$ is $\epsilon$-correctable for $\chan$, i.e. $\|\mathcal R \circ \cchan - \mathcal P_{\mathcal A}\|_\diamond \le \epsilon$. The by using again the result of~\cite{kretschmann08} we have that $\widehat {R \circ \chan}$ is within $2\sqrt \epsilon$ to some channel $\mathcal M$ complementary to $\mathcal P_{\mathcal A}$. But since $\mathcal A$ is obviously correctable for the channel $\mathcal P_{\mathcal A}$, we know by the condition for exact correction (Equ.~\ref{gkl3}) that $\mathcal M = \mathcal M \circ \mathcal P_{\mathcal A'}$:
\[
\| \widehat {R \circ \chan} - \mathcal M \circ \mathcal P_{\mathcal A'} \|_{\diamond} \le 2 \sqrt{\epsilon}.
\]
Note that we can define a dilation of $\mathcal R \circ \chan$ by the isometry $V = (V_{\mathcal R} \otimes \one) V_\chan $ where $V_{\mathcal R}$ (resp. $V_\chan$) defines a dilation of $\mathcal R$ (resp. $\chan$). In this product, the input of $V_{\mathcal R}$ is the output of $\chan$. Hence the corresponding channel complementary to $\mathcal R \circ \chan$ has two outputs, one from $V_\chan$ and one from $V_{\mathcal R}$. If we trace out the output of $V_{\mathcal R}$ we obtain a channel complementary to $\chan$, which we will call $\cchan$. Applying the same partial trace on $\mathcal M \circ \mathcal P_{\mathcal A'}$ yields a channel $\mathcal M' = \mathcal M' \circ \mathcal P_{\mathcal A'}$. Given that a partial trace cannot increase the diamond norm, we obtain
\[
\| \cchan - \mathcal M'\|_\diamond = \| \cchan - \mathcal M' \circ \mathcal P_{\mathcal A'}\|_\diamond \le 2\sqrt{\epsilon}
\]
Hence also,
\[
\| \cchan  \circ \mathcal P_{\mathcal A'} - \mathcal M' \|_\diamond \le \| \cchan - \mathcal M'\|_\diamond \|\mathcal P_{\mathcal A'}\|_{\diamond} \le 2\sqrt{\epsilon}.
\]
If we use these two inequalities together we obtain
\[
\| \cchan - \cchan \circ \mathcal P_{\mathcal A'}\|_\diamond \le \| \cchan - \mathcal M'\|_\diamond + \| \mathcal M' - \cchan \circ \mathcal P_{\mathcal A'}\|_\diamond \le 4 \sqrt{\epsilon}.
\]
\qed

\vspace{0.2in}

As an example let us show how the estimate (Equ. ~\ref{estim}) looks like if we want to approximately correct a subspace, i.e. when the algebra $\mathcal A$ consists of the set of all operators acting on a code space $\Hil_C$, which we write $\mathcal A = \mathcal B(\Hil_C)$. Let $V$ be the isometry embedding $\Hil_C$ into the physical Hilbert space $\Hil$. We also write the encoding channel as $\mathcal E(\rho) = V \rho V^\dagger$. In this case, the commutant $\mathcal A'$ is the trivial algebra containing only multiples of the identity on $\Hil_C$. Hence the corresponding projector is
\[
\mathcal P_{\mathcal A}(\rho) = \tr(\rho) \frac{\one}{\tr{\one}}.
\]

Let $d$ be the dimension of the code Hilbert space $\Hil_C$, and
\[
\lambda_{ij} := \bra{i}\widehat {\chan \circ \mathcal E}(\one/d)   \ket{j} = \frac{1}{d} \tr(V^\dagger E_i^\dagger E_j V)
\]
then direct computation shows that our estimate of the optimal recovery error is 
\begin{equation}
\label{akl}
\begin{split}
\delta_{\mathcal B(\Hil_C)}(\chan \circ \mathcal E) 
&= \sup_{\| B \| \le 1} \| \sum_{ij} (V^\dagger E_i^\dagger E_j V - \lambda_{ij} \one) \otimes  B_{ij} \|
\end{split}
\end{equation}
where $B_{ij}$ are blocks of $B$, i.e. $B = \sum_{ij} \ketbra i j \otimes  B_{ij}$.  
We see that the exact Knill-Laflamme conditions put this quantity to zero by imposing $V^\dagger E_i^\dagger E_j V - \lambda_{ij} \one = 0$ for all $i$, $j$.


\subsection{Condition on individual operators}


In the case of exact correction, the form of the condition expressed as a commutator (Equ.~\ref{gkl2}) is fundamental because it shows how different correctable algebras are related: namely that they are all in fact part of a largest correctable algebra. Here we show that a form of this condition still holds in the approximate case, however the consequences are weaker. 


The following lemma will allow us to make this generalization. 
\begin{lemma}
\label{thelemma}
Let $\mathcal A$ be a $\dagger$-algebra and $B$ any operator with $\|B\| \le 1$, then
\[
\| B - \mathcal P_{\mathcal A'} (B) \| \le \sup_{A \in \mathcal A, \|A\|\le 1} \|[A, B]\| \le 2\| B - \mathcal P_{\mathcal A'} (B) \|
\]
\end{lemma}
\Prf
The upper bound is straightforward:
\[
\begin{split}
\| [A, B]\| &= \|AB - A \mathcal P_{\mathcal A'} (B) +  \mathcal P_{\mathcal A'} (B) A - BA  \|\\
 &\le \|A\|\|B - \mathcal P_{\mathcal A'} (B)\| +  \|\mathcal P_{\mathcal A'} (B)  - B\|\| A \|\\
&\le 2 \| B - \mathcal P_{\mathcal A'} (B) \|.
\end{split}
\]

For the lower bound, note that the set of unitary operators in $\mathcal A$ forms a group with Haar measure $\mu$. 
We assume that the measure is normalized to one. 
Note that the projector $\mathcal P_{\mathcal A'}$ can be computed by averaging over this group:
\[
\mathcal P_{\mathcal A'} (B) = \int d\mu(U) U^\dagger B U
\]
for all $B$.
Indeed, it is clear that $\mathcal P_{\mathcal A'} = \mathcal P_{\mathcal A'}^\dagger$, and the fact that
\begin{equation}
\label{pinv}
U^\dagger \mathcal P_{\mathcal A'} (B) U =  \mathcal P_{\mathcal A'} (B)
\end{equation}
implies $\mathcal P_{\mathcal A'}^2 = \mathcal P_{\mathcal A'}$.
In addition Equ.~\ref{pinv} also implies, 
\(
[ \mathcal P_{\mathcal A'} (B), U ] = 0
\)
for all $U \in \mathcal A$, which implies $\mathcal P_{\mathcal A'} (B) \in \mathcal A'$ for all $B$ since the unitary operators span the algebra. But also, since all the unitary operators integrated over are in $\mathcal A$, it is clear that $\mathcal P_{\mathcal A'} (A)  = A$ for all $A \in \mathcal A'$. 

Using this expression for $\mathcal P_{\mathcal A'}$, we have
\[
\begin{split}
\| B - \mathcal P_{\mathcal A'} (B) \| &\le \int d\mu(U) \|U^\dagger U B - U^\dagger B U \|\\
&\le \int d\mu(U) \|U^\dagger\| \| [U, B ]\|\\
&\le \sup_{A \in \mathcal A, \|A\| \le 1} \|[A, B]\|
\end{split}
\]
\qed

We can now combine this lemma with Theorem~\ref{mainthm} to obtain the following condition for approximate correctability:
\begin{theorem}
\label{secthm}
If an algebra $\mathcal A$ is $\frac{1}{8}\epsilon^2$-correctable then all its elements $A \in \mathcal A$ with $\|A\| \le 1$ must approximately commute with all the observables preserved in the environment, i.e.
\[
\| [A \otimes \one, (\cchanh \otimes \id)(B)] \| \le \epsilon 
\]
for all operators $\|B\| \le 1$, where $\one$ and $\id$ act on a Hilbert-space of dimension equal to that of the source of $\chan$. Conversely, this condition guarantees that $\mathcal A$ is $2\sqrt{\epsilon}$-correctable. 
\end{theorem}
\proof
The estimate $\delta_{\mathcal A}(\chan)$ defined in Equ.~\ref{estim} can be expressed in terms of the CB norm distance between the dual channels as
\[
\delta_{\mathcal A}(\chan) = \| \cchanh - \mathcal P_{\mathcal A'} \circ \cchanh  \|_{cb}.
\]
Theorem~\ref{mainthm} then implies that if an algebra $\mathcal A$ is $\frac{1}{8}\epsilon^2$-correctable, then for all $\|B\| \le 1$,
\[
\| (\cchanh \otimes \id)(B) - (\mathcal P_{\mathcal A'} \otimes \id)(\cchanh \otimes \id)(B)\| \le \frac{1}{2}\epsilon.
\]
Since $\mathcal P_{\mathcal A'} \otimes \id$ is just the projector on the algebra $\mathcal A \otimes B(\Hil)$, Lemma~\ref{thelemma} implies that for all $A \in \mathcal A$,
\[
\| [A \otimes \one, (\cchanh \otimes \id)(B)] \| \le \epsilon.
\]
Reciprocally, following the same steps in reverse, this condition implies via Lemma~\ref{thelemma} that $\delta_{\mathcal A}(\chan) \le \epsilon$ which then implies via Theorem~\ref{mainthm} that $\mathcal A$ is $2\sqrt{\epsilon}$ correctable.
\qed

We see that contrary to the exact case, the approximately correctable observables must not only approximately commute with the observables preserved by the complementary channel, but also with the observables preserved by its trivial extension on a larger space. 


Also, unlike in the exact case, this condition does not guarantee that it is sufficient to test the commutativity condition on generators of the algebra. We can only rely on the convexity of the approximate condition. For instance this bound degrades with the number of products taken. If the norm one operators $A_i$, $i=1,\dots,n$ satisfy $\| [A_i \otimes \one, (\cchanh \otimes \id)(B)] \| \le \epsilon$, then we can only guarantee that $\| [A_1A_2\cdots A_n \otimes \one, (\cchanh \otimes \id)(B)] \| \le n \epsilon$.






\section{Outlook}

In the exact case we know that there is only one maximal set of simultaneously correctable observables; the commutant of the operators $E_i^\dagger E_j$. We hope that the results presented in the last section can help understand the structure of---and the relation between---the sets of simultaneously approximately correctable observables, for a given error $\epsilon$.

We focused here on the diamond norm distance between the corrected channel $\mathcal R \circ \chan$ and a target channel $\mathcal P_{\mathcal A}$ because it allowed us to obtain a generalization of the condition expressed in terms of the commutation relation (Equ. ~\ref{gkl2}). However, the same technique yields a much tighter estimate of the worst case entanglement fidelity. This will be analyzed in a separate article in which we will also address the problem of finding a good approximate correction channel.




\vspace{0.4in}

\section*{Acknowledgements}
Part of this work was done at the workshop QIP 2009. 
The author is grateful to the participants of the workshop TQC 2009
for feedback. The Centre for Quantum Technologies is funded by the Singapore Ministry of Education and the National Research Foundation as part of the
Research Centres of Excellence programme.

\bibliographystyle{hunsrt}
\bibliography{pig} 

\end{document}